\documentclass[prb,twocolumn,preprintnumbers,amsmath,amssymb]{revtex4}
\usepackage{graphicx}
\usepackage{epsfig}
\usepackage{amsmath,amscd}

\def\rl{{\bf r}_{L}}  
\def\vl{{\bf v}_{L}}  
\def\vs{{\bf v}_{s}}  
\def\vn{{\bf v}_{n}}  
\def\Js{{\bf J}_{s}}  
\def\Jn{{\bf J}_{n}}  
  
\def\Ep{{\bf E}}  
\def\Eloc{{\Ep}}  
\def\Bloc{{\bf B}}  

\def\secref#1{\ref{#1}}

\def\olcite#1{[\onlinecite{#1}]}

\def\D{{\bf D}}
\def\Ups{{\bf \Upsilon}}
\def\DbD{\overleftarrow{\overline{\D}} - \D}

\def\nn{\nonumber}
\def\del{\partial}
\begin{document}
\title{Extended Josephson relation and Abrikosov lattice deformation}
\author{Peter Matlock}
\email{pwm@induulge.net}
\affiliation{Research Department, Universal Analytics Inc., Airdrie, AB, Canada}
\begin{abstract}
\noindent
From the point of view of time-dependent Ginzburg Landau (TDGL)
theory, a Josephson-like relation is derived for an Abrikosov vortex
lattice accelerated and deformed by applied fields. Beginning with a
review of the Josephson Relation derived from the two ingredients of a
lattice-kinematics assumption in TDGL theory and gauge invariance, we
extend the construction to accommodate a time-dependent applied
magnetic field, a floating-kernel formulation of normal current, and
finally lattice deformation due to the electric field and inertial
effects of vortex-lattice motion. The resulting Josephson-like relation,
which we call an Extended Josephson Relation, applies to a much wider
set of experimental conditions than the original Josephson Relation,
and is explicitly compatible with the considerations of TDGL theory.
\end{abstract}
\maketitle

\section{Introduction}

The well-known Josephson Relation\cite{Josephson} (JR) relates applied
electric and magnetic fields to motion of the Abrikosov vortex lattice
in a thin superconducting sample.  Taking a simple form,
\begin{equation}
\label{JR}
\langle \Ep \rangle = -\frac{1}{c} \vl \times \langle \Bloc \rangle,
\end{equation}
the JR is easily understood intuitively via arguments involving the 
Faraday Law, and motion of magnetic flux, and thought to 
represent an approximation of general applicability, therefore being widely
employed by experimentalists. It is thus time-tested and, though not of a
fundamental theoretical nature, is an essential and useful tool.
Here, $\Eloc$ and $\Bloc$ will denote the local electric 
and magnetic fields, and $\langle\cdots\rangle$ denotes an averaging in 
space.

Relation \eqref{JR} has been known for decades, and does also stand up
to more modern considerations in the context of time-dependent
Ginzburg-Landau (TDGL) theory. Indeed, the JR is derived by
Kopnin\cite{Kopnin}, using some simple assumptions about the rigidity
of the vortex-lattice motion. 
In this analysis, the premise is that the vortex lattice already 
represents some solution to TDGL theory, and its motion must remain 
compatible with that theory; therefore there is an implicit assumption
that the vortex lattice is triangular.

More recently, a derivation along the
same lines but accommodating inertial effects of the condensate
has been exhibited\cite{LipavskyIJR}, which reproduces from the point
of view of TDGL theory an Inertial Josephson Relation (IJR), and which had
been obtained much earlier in a hydrodynamic
context\cite{Abrikosov,Hocquet}.  The IJR contains a term 
reflecting the inertia of the condensate itself, and
is thus an extension of the JR, applicable at high
frequencies.  We shall review this result below in section
\secref{secJRIJR}, obtaining the IJR using the requirement of gauge
invariance, combined with simple ingredients of rigid lattice motion
and the current obtained from TDGL theory.

In the present work we shall discuss ways in which this analysis can
naturally be taken further, producing what we shall call an Extended
Josephson Relation. The Inertial Josephson relation mentioned above
represents one extension of the Josephson relation to accommodate
condensate inertia. By Extended Josephson Relation, we shall mean
a relation further extended to include the effects discussed below.

The starting point is chosen to be the Floating-Kernel\cite{LL08,LL09}
(FK) version of the TDGL equation, which contains the normal current;
so named because it can be understood by shifting to the reference
frame `floating' with this normal current.

Not only can a normal-current correction be included, but it is not
necessary to work with the assumption of a rigid Abrikosov vortex
lattice.  Of course, should the (average) magnetic field change with
time, so must the density of unit-flux vortices.

Finally, since the electric field and also the inertia of the vortex
lattice itself provide anisotropic stimuli in the plane of the vortex
lattice, one ought consider deformations of the lattice which include
not only the density fluctuations mentioned above, but a group of
geometrical deformations which represent a response to these
anisotropies.

An Extended Josephson Relation is calculated in section
\secref{secEJR} which accommodates all these novel ingredients.  The
result can be considered an extension of the Josephson Relation (and
Inertial Josephson Relation) which can describe density fluctuations
of the vortex lattice which allow for a time-varying magnetic field,
and anisotropic deformations of the vortex lattice from the electric
field and vortex acceleration.

\section{Notation, conventions and basics}
We consider the usual configuration of a planar sample in the $x$-$y$
plane, with $\Ep \perp \hat{z}$ and vortices produced by $\Bloc
\parallel \hat{z}$.  We will make use of the usual definitions $\Ep =
- \frac{1}{c} \del_t {\bf A} - \nabla \phi$ and $\Bloc = \nabla
\times {\bf A}$.

We write a gauge transformation as
\begin{eqnarray}
\label{gauge1}
{\bf A} &\rightarrow&  {\bf A} + c \nabla\theta \\
\label{gauge2}
\phi &\rightarrow&  \phi -  \del_t \theta \\
\label{gauge3}
\psi &\rightarrow& e^{ie^*\theta/\hbar}\psi ,
\end{eqnarray}
where $\theta$ is a function of space and time.
Our calculation will be gauge-invariant; we will make no gauge choice.
This allows us to use the principle of gauge covariance.

\section{IJR from TDGL}
\label{_IJRfromTDGL}
\label{secJRIJR}
Within this section, we carefully extract the Inertial Josephson
Relation from the TDGL equation.  We follow closely the derivation of
the Josephson Relation given by Kopnin \cite{Kopnin} or the IJR given
in \olcite{LipavskyIJR}; we pay special attention to gauge invariance,
in anticipation of extending the IJR later.

The TDGL equation is 
\begin{eqnarray}
\label{_TDGL}
&&\frac{-\hbar^2}{2m^*} {\D}^2 \psi + \alpha \psi + \beta \psi |\psi|^2 \nn\\
&=&\frac{-\Gamma}{\sqrt{1+C^2|\psi|^2}} \big( \del_t + \frac{i}{\hbar}e^*\phi + \frac{c^2}{2}\del_t |\psi|^2 \big)\psi
\end{eqnarray}
where the gauge-covariant derivative is 
\begin{equation}
{\D}\psi := (\nabla - \frac{ie^*}{\hbar c} {\bf A} )\psi 
.
\end{equation}
$\Js$ resulting from \eqref{_TDGL} is given by variation of ${\bf A}$;
\begin{equation}
\label{_superJ}
\Js = \frac{i\hbar e^*}{2 m^*} \bar{\psi} \big( \DbD \big)  \psi 
.
\end{equation}

Now we must formulate the assumption of rigid motion of the vortex lattice.
We define by $\psi_0$ the configuration at time zero, $\psi_0({\bf r}) := \psi({\bf r},0)$.
With the displacement of the vortex lattice given by $\rl(t)$, we thus require
\begin{equation}
\label{shiftpsi2}
|\psi({\bf r},t)|^2 = |\psi_0({\bf r}-\rl(t))|^2
\end{equation}
so that for some \emph{pure-gauge} function $\omega$ we have
\begin{equation}
\label{shiftpsi}
\psi({\bf r},t) = e^{-i\omega({\bf r},t)} \psi_0({\bf r}-\rl(t)).
\end{equation}
This equation must be gauge-invariant; transforming both 
sides using $\theta({\bf r},t)$ (see \eqref{gauge1} -- \eqref{gauge3}), 
we see that $\omega$ must have the gauge transformation law
\begin{equation}
\label{omegagauge}
\omega({\bf r},t) \rightarrow \omega({\bf r},t) + \frac{e^*}{\hbar}\big( \theta({\bf r} - \rl(t),0) - \theta({\bf r},t)\big).
\end{equation}
Whatever physical assumption we specify, in the present case equation \eqref{shiftpsi},
must be independent of any gauge choice. Although we do not at this stage know 
the form of the function $\omega$, we do know how it must transform, 
and we shall make use of the transformation law \eqref{omegagauge} in what follows.

Writing $\vl = \del_t \rl$, we use \eqref{shiftpsi} to calculate
\begin{eqnarray}
\del_t \big[ \bar{\psi} \big( \DbD \big)  \psi \big] 
&=& -\vl\cdot\nabla \big[ \bar{\psi} \big( \DbD  \big)  \psi \big] \\
&+& 2i|\psi|^2 (\vl\cdot\nabla + \del_t ) \big( \nabla\omega + \frac{e^*}{\hbar c} {\bf A}\big) .\nn
\end{eqnarray}
Equation \eqref{omegagauge} implies that $(\vl\cdot\nabla + \del_t ) \omega$ gauge-transforms as
\begin{equation}
\label{omegagauge2}
 (\vl\cdot\nabla + \del_t ) \omega \rightarrow  (\vl\cdot\nabla + \del_t ) \omega 
- \frac{e^*}{\hbar} (\vl\cdot\nabla + \del_t ) \theta,
\end{equation}
which determines 
\begin{equation}
 (\vl\cdot\nabla + \del_t ) \omega  = \frac{e^*}{\hbar}\big( \phi - \frac{1}{c} \vl\cdot{\bf A} \big).
\end{equation}
Using 
\begin{equation}
\label{_covvs}
\vs := \frac{\Js}{ e^* |\psi|^2}
\end{equation}
 we then obtain
\begin{equation}
\del_t \vs = -\vl \cdot \nabla \vs + \frac{e^*}{m^*} \big[ \frac{1}{c} \nabla(\vl\cdot{\bf A})
- \frac{1}{c}\vl\cdot\nabla {\bf A} + \Ep  \big]
.
\end{equation}
Finally, use the identity $\nabla ( {\bf v} \cdot {\bf V}) - {\bf v}
\cdot \nabla {\bf V} = {\bf v} \times \nabla \times {\bf V}$ for
constant ${\bf v}$ to write
\begin{equation}
\del_t \vs = -\vl \cdot \nabla \vs + \frac{e^*}{m^*c} \vl\times\Bloc + \frac{e^*}{m^*} \Ep
.
\end{equation}
Note that this tells us the field $\vs$ does not move with the lattice; otherwise we would have $(\del_t + \vl\cdot\nabla)\vs=0$.

We may now take the unit-cell average. Since $\vs$ is periodic, the first term does not contribute and
\begin{equation}
\label{_IJR}
\del_t \langle \vs \rangle = \frac{e^*}{m^*}\langle\Ep\rangle + \frac{e^*}{m^*c} \vl \times \langle B\rangle 
.
\end{equation}
This is the IJR as presented in \olcite{LipavskyIJR}.
It can be thought of as the consequence of describing a 
rigidly moving vortex lattice using TDGL theory, and includes
the inertial term $\del_t \langle \vs \rangle $, absent 
from the original Josephson Relation.

\section{Floating-Kernel TDGL}
\label{SecFKTDGL}
Our starting point will not be the TDGL equation as
presented in \eqref{_TDGL}, but a version of TDGL supplemented by a
floating kernel (FK) term,\cite{LL08,LL09}
\begin{eqnarray}
\label{_TDGLfk}
&&\frac{1}{2m^*} \big( -i\hbar\nabla - \frac{e^*}{c} {\bf A} - \frac{m^*}{en} \Jn \big)^2 \psi + \alpha \psi + \beta \psi |\psi|^2 \nn\\
&=&\frac{-\Gamma}{\sqrt{1+C^2|\psi|^2}} \big( \del_t \frac{i}{\hbar}e^*\phi + \frac{c^2}{2}\del_t |\psi|^2 \big)\psi
.
\end{eqnarray}
The supercurrent is defined though variation of ${\bf A}$,
\begin{equation}
\label{_fksuperJ}
\Js = \frac{e^*}{m^*} \bar{\psi} 
\big( \frac{i\hbar}{2} ( \DbD )  - \frac{m^*}{en}\Jn \big)
 \psi 
.
\end{equation}
In the spirit of a two-fluid model of superconductivity, we write
\begin{equation}
\label{Jsvsvn}
\Js = e^*|\psi|^2(\vs-\vn)
\end{equation}
and write the total current as a sum
\begin{eqnarray}
{\bf J} &:=& \Js + \Jn \\
&=& e^*n_s\vs + en_n \vn
\end{eqnarray}
where $\vn := \Jn/en$, $\Jn = \sigma_n {\bf E}$ and $n = n_n + 2 n_s$.

In the following section, we do not attempt to solve TDGL theory to
determine the density or deformation dynamics of the vortex lattice,
just as in the previous section there was no attempt to solve TDGL to
determine rigid motion of the vortex lattice. Instead, we allow that
it may happen and study the consequences. In fact, rigid lattice
motion may have been too strict an assumption even in the case of no
normal current; there is no \emph{a priori} reason why the magnetic 
field must always remain constant, and the lattice rigid.

\section{Extended Josephson relation}
\label{secEJR}
In this section we allow much greater freedom for the moving vortex
lattice; in particular, we allow the magnitude of $\Bloc$ to change
with time, so that the vortex lattice density is time-dependent.
Further, we allow the shape of the vortex lattice to undergo a global 
time-dependent deformation, in response to anisotropic stimuli.

We shall generalise the rigid-motion requirement \eqref{shiftpsi} to
\begin{equation}
\label{psiomegatilde}
\psi({\bf r},t) = e^{-i\tilde{\omega}({\bf r},t)}\psi(\lambda(t)\Sigma(t)({\bf r} - \rl(t)),0)
.
\end{equation}
Here $\lambda(t)$ is a dynamic scaling factor for the vortex lattice.
We can expect such a simple dynamic scaling to be valid for a 
$\Bloc$ field which does not vary excessively or too quickly with time.
We require $\rl(0)=0$ and $\lambda(0)=1$, and define $B_0$ as the 
magnitude of the average magnetic field when the vortex lattice density is $\lambda=1$,
therefore $B(t) \equiv \langle |{\bf B}(t)| \rangle = B_0 \lambda^2(t)$.

$\Sigma(t)$ is a two-dimensional dynamic deformation matrix for the
vortex lattice. Since we have accommodated scaling with $\lambda$, we
shall require that $\Sigma$ be an element of $SL(2,\mathbb{R})$, and
we will write 
\begin{equation}
\label{deformS}
\Sigma(t)=e^{S(t)}.
\end{equation}
We shall leave discussion of the matrix $S$ to the following section.

Beginning with the vortex lattice \eqref{psiomegatilde} and the TDGL
equation with floating kernel \eqref{_TDGLfk}, we proceed as in the
previous section.
Using equation \eqref{psiomegatilde} we calculate
\begin{equation}
\label{dtpsi}
\del_t \psi = \Ups\cdot\nabla \psi + i \Omega \psi
\end{equation}
where
\begin{eqnarray}
\label{_Upsilon}
\Ups &=& \lambda^{-1}\Sigma^{-1} \del_t \big[ \lambda(t)\Sigma(t)({\bf r} - \rl(t)) \big] ,\\
\label{_Omega}
\Omega &=& \Ups\cdot\nabla\tilde{\omega} - \del_t \tilde{\omega},
\end{eqnarray}
so that
\begin{eqnarray}
\del_t \D \psi &=& (\Ups\cdot\nabla + i\Omega ) D \psi + \nabla(\Ups\cdot\nabla+i\Omega) \psi \nn\\
 &+& \frac{ie^*}{\hbar c} (\Ups\cdot\nabla - \del_t ){\bf A} \psi
\end{eqnarray}
and
\begin{eqnarray}
\del_t \frac{\bar{\psi} (\DbD)\psi }{|\psi|^2} &=& [ \Ups\cdot\nabla + \nabla\Ups\cdot ] \frac{\bar{\psi} (\DbD)\psi }{|\psi|^2} \nn\\
&+& \frac{2ie^*}{\hbar c} [  \del_t - \Ups\cdot\nabla - \nabla\Ups\cdot ]{\bf A}  \nn\\
&-& 2i\nabla\Omega .
\end{eqnarray}
This allows us to use \eqref{Jsvsvn} and evaluate
\begin{eqnarray}
\label{dtvsvn}
\del_t (\vs - \vn) &=&  (\Ups\cdot\nabla + \nabla\Ups\cdot ) (\vs - \vn) \nn\\
&+&\frac{e^*}{m^*c} \big[ \nabla(\Ups\cdot{\bf A}) - \Ups\times{\bf B} \big] \nn\\
&+& \frac{e^*}{m^*} \big[ {\bf E} + \nabla\phi + \frac{\hbar}{e^*} \nabla\Omega \big] \nn\\
&-&\frac{2}{n}\del_t\Jn
\end{eqnarray}
where this time we have used the identity 
\begin{equation}
 \nabla({\bf v}\cdot{\bf  V}) - {\bf v}\cdot\nabla{\bf V} = 
{\bf v}\times \nabla\times {\bf V} + \nabla{\bf v}\cdot{\bf V}.
\end{equation}
Now, although $\Ups$ is gauge-invariant, it may be seen from 
\eqref{_Omega} or \eqref{dtpsi} that $\Omega$ must transform as
\begin{equation}
\Omega \rightarrow \Omega - \frac{e^*}{\hbar}[ \Ups\cdot\nabla\theta - \del_t\theta ]
\end{equation}
which determines
\begin{equation}
\Omega = - \frac{e^*}{\hbar}[ \frac{1}{c} \Ups\cdot{\bf A} + \phi ].
\end{equation}
\def\rrl{({\bf r}-\rl)}
Substituting $\Omega$ and $\Ups$ into \eqref{dtvsvn}, we find
\begin{eqnarray}
\label{dtvsvnsimp}
\del_t (\vs - \vn) &=&  \big[ \frac{\del_t\lambda}{\lambda}\rrl\cdot\nabla - \vl\cdot\nabla \big](\vs-\vn)\nn\\
&+& \big[  \rrl\del_tS^T\nabla  - \frac{\del_t\lambda}{\lambda}  + \del_tS^T \big] (\vs-\vn) \nn\\
 &+& \frac{e^*}{m^*}(1-\tau\del_t){\bf E} - \frac{e^*}{m^*c} 
\big[ \frac{\del_t\lambda}{\lambda}\rrl \nn\\
&+& \del_tS \rrl - \vl \big]\times{\bf B}
\end{eqnarray}
where $\tau = m^*\sigma_n/2e^2n$. Finally we take the spatial average and find
\begin{eqnarray}
\label{EJRS}
\del_t \langle\vs - \vn\rangle &=& \frac{e^*}{m^*}(1-\tau\del_t) \langle{\bf E}\rangle 
                                   + \frac{e^*}{m^*c}\vl\times\langle{\bf B}\rangle \nn\\
&+& \big[ \del_tS^T - \frac{\del_t B}{2B} \big] \big( \langle\vs-\vn\rangle + \frac{2}{n}\sigma_n \langle{\bf E}\rangle \big) \nn\\
&+& \frac{e^*}{m^*c}\big[ \frac{\del_t B}{2B} \rl + \del_tS\rl \big]\times\langle{\bf B}\rangle
\end{eqnarray}
where we have substituted for $\lambda$. We have used that $\vs$ is
periodic and $\vn$ is uniform, and for the sake of simplicity we have
taken the origin ${\bf r}=0$ to be at the centroid of the sample.
Equation \eqref{EJRS} is our Extended Josephson Relation (EJR), the main
result of this paper, and we pause to comment.  The first line of
\eqref{EJRS} is simply the Inertial Josephson Relation of
\eqref{_IJR}, with the addition of the $\tau$ term accounting for the
floating-kernel normal current contribution. The remaining corrections
consist of parts which depend on the time derivative of the magnetic
field, and parts which depend on geometrical lattice deformation; when
deformation is absent, $S=0$.

In the following section we complete the expression of the EJR
\eqref{EJRS} by showing the explicit parametrisation of the
deformation matrix $S$,

\section{Lattice deformation}
\label{LatDef}

\def\TDM#1#2#3#4{\left[ \begin{tabular}{cc}$ #1$ &$ #2$ \\$ #3$ & $#4$ \end{tabular}\right]}
A two-dimensional space may be expanded by a scale factor $e^\sigma$ in the $x$ direction,
and $e^{-\sigma}$ in the $y$ direction by the matrix
\begin{equation}
\Sigma(0,\sigma) = \TDM{e^\sigma}{0}{0}{e^{-\sigma}}
\end{equation}
which has unit determinant.
Rotating so that the expansion is along a line at angle $\vartheta$ and the contraction 
perpendicular,
\def\bmu{{\boldsymbol \mu}}
\def\bnu{{\boldsymbol \nu}}
\begin{eqnarray}
\Sigma(\vartheta,\sigma) &=& R(-\vartheta) \Sigma(0,\sigma) R(\vartheta) \nn\\
&=& \exp\left\{ \sigma \TDM{1-2\sin^2\vartheta}{2\sin\vartheta\cos\vartheta}{2\sin\vartheta\cos\vartheta}{1-2\cos^2\vartheta} \right\} \nn\\
&:=& \exp S_{\bmu}.
\end{eqnarray}
Here we have parametrised the deformation by ${\bmu} := (\sigma \cos \vartheta , \sigma \sin \vartheta) $.
This set of deformations is not closed; composition can produce rotations.
When combined with rotations all elements of $SL(2,{\mathbb R})$ can be constructed.
Some identities are $\Sigma(\vartheta,\sigma)^{-1}=\Sigma(\vartheta+\pi/2,\sigma)=\Sigma(\vartheta,-\sigma)$,
$\Sigma(\vartheta,0)=1$ and $\Sigma(\vartheta,\sigma_2)\Sigma(\vartheta,\sigma_1)=\Sigma(\vartheta,\sigma_1+\sigma_2)$.

If we define convenient matrices which depend on the deformation parameter $\bmu$,
\begin{equation}
\label{mandn}
m_{\bmu} := \TDM{\mu_x}{\mu_y}{\mu_y}{-\mu_x}\quad \textup{and}\quad n_{\bmu} := \TDM{\mu_x}{\mu_y}{-\mu_y}{\mu_x},
\end{equation}
we may write $ S_{\bmu} = {m_{\bmu} n_{\bmu}} / {\mu} $.
For a given time-dependent deformation parameter ${\bmu}(t)$, we may calculate
\begin{equation}
\label{dtSmu}
\del_t S_{{\bmu}(t)} = \frac{m_{\bmu}}{\mu} \big[  2n_{{\bmu}'} - \frac{{\bmu}\cdot{\bmu}'}{\mu^2} n_{\bmu} \big].
\end{equation}

Now, let us recall the $S$ matrix introduced in section \ref{secEJR},
equation \eqref{deformS}.  Given some deformations parametrised by
$\bmu$, $\bnu$, $\dots$, we would set $S = S_{\bmu} + S_{\bnu} +
\cdots$.  In fact, it is possible immediately to write down several
plausible examples of physical sources of deformation.

Let us anticipate that the application of an electric field (in the
plane), or a time-derivative of this field may tend to deform the
vortex lattice. We can accommodate this with a deformation parameter 
\begin{equation}
\label{def_field}
\bnu := E_0 {\bf E} + E_1 \del_t {\bf E}.
\end{equation}
Another possibility is that the global inertia or acceleration of 
the vortex lattice would coincide with a deformation; in this 
case,
\begin{equation}
\label{def_iner}
\bmu := v_0 \vl + v_1 \del_t\vl.
\end{equation}
Obviously, there is an approximation involved here in that our 
consideration is limited to a global deformation.

Adding the above two deformation contributions together, 
\begin{equation}
S(t) = S_{\bnu} + S_{\bmu}.
\end{equation}
$\del_tS(t)$ may be calculated for each of the two terms by using \eqref{dtSmu};
\begin{eqnarray}
\del_t S(t) &=&
 \frac{m_{\bmu}}{\mu} \big[  2n_{{\bmu}'} - \frac{{\bmu}\cdot{\bmu}'}{\mu^2} n_{\bmu} \big] \nn\\
&+&
 \frac{m_{\bnu}}{\nu} \big[  2n_{{\bnu}'} - \frac{{\bnu}\cdot{\bnu}'}{\nu^2} n_{\bnu} \big].
\end{eqnarray}
Upon substitution of $\del_tS$ and $\del_tS^T$ the Extended Josephson 
Relation \eqref{EJRS} will contain the parameters $E_0$, $E_1$, $v_0$ and $v_1$.
In principle, these parameters could be fit to experimental data,
characterising response beyond the usual JR or IJR.

\section{Conclusions}
\label{secConc}

As we mentioned in the introduction, the Josephson Relation, though
simple in form, is often used by experimentalists to understand the
motion of vortices in the presence of an applied field. It has been
shown in the literature how to recover the Josephson
Relation\cite{Kopnin} and a form valid at higher frequencies, the
Inertial Josephson Relation,\cite{LipavskyIJR} in the context of TDGL
theory. We have taken this technique further, using the assumption of
a vortex lattice solution to TDGL theory, to extend the Josephson
Relation to a form covering a floating-kernel formulation of TDGL with
a normal current in the spirit of a two-fluid model. Additionally, we
have shown how to include effects of time-varying magnetic flux, and
also global vortex-lattice deformations; our considerations are
expected to be valid for small changes in magnetic field and small
lattice deformations.  This is not due to any approximation in the
calculation itself, rather it is due to the implicit assumption of the
characteristics of the vortex lattice dynamics, equation
\eqref{psiomegatilde}, which nevertheless represents a far weaker
assumption than that of perfectly rigid global motion
\eqref{shiftpsi2}.

\section*{Acknowledgements}
The author is grateful to P.-J.~Lin and P.~Lipavsk\'y for fruitful discussions 
regarding Josephson Relations.


\begin{thebibliography}{99}
\bibitem{Josephson} B.~D.~Josephson, Phys.~Lett.~16, 242 (1965)
\bibitem{Kopnin} N.~B.~Kopnin, \emph{Theory of Nonequilibrium Superconductivity}, Claredon Press, Oxford (2001)
\bibitem{LipavskyIJR} P.-J.~Lin, P.~Lipavsk\'y and Peter~Matlock, Phys.~Lett.~A 376 (2012) pp. 883-885
\bibitem{Abrikosov} A.~A.~Abrikosov, M.~P.~Kemoklidze, and I.~M.~Khalatnikov, Sov.~Phys.~JETP 21, 506 (1965)
\bibitem{Hocquet} T.~Hocquet, P.~Mathieu, and Y.~Simon, Phys.~Rev.~B 46, 1061 (1992)
\bibitem{LL08} P.-J.~Lin and P.~Lipavsk\'y, Phys.~Rev.~B 77, 144505 (2008)
\bibitem{LL09} P.-J.~Lin and P.~Lipavsk\'y, Phys.~Rev.~B 80, 212506 (2009)
\end{thebibliography}
\end{document}